# Modelling Air Pollution Crises Using Multi-agent Simulation

Sabri Ghazi
Computer Science department
University Badji Mokhtar,
PO-Box 12, 23000,
Annaba, Algeria
ghazi@labged.net

Julie Dugdale
University Grenoble Alps.
LIG. France.
University of Agder, Agder
Norway
Julie.Dugdale@imag.fr

Tarek Khadir
Laboratoire de gestion
électronique du document
Computer Science department
University Badji Mokhtar,
PO-Box 12, 23000,
Annaba, Algeria
Khadir@labged.net

**Abstract**
*This paper describes an agent based approach for simulating the control of an air pollution crisis. A Gaussian Plum air pollution dispersion model (GPD) is combined with an Artificial Neural Network (ANN) to predict the concentration levels of three different air pollutants. The two models (GPM and ANN) are integrated with a MAS (multi-agent system). The MAS models pollutant sources controllers and air pollution monitoring agencies as software agents. The population of agents cooperates with each other in order to reduce their emissions and control the air pollution. Leaks or natural sources of pollution are modelled as uncontrolled sources. A cooperation strategy is simulated and its impact on air pollution evolution is assessed and compared. The simulation scenario is built using data about Annaba (a city in North-East Algeria). The simulation helps to compare and assess the efficiency of policies to control air pollution during crises, and takes in to account uncontrolled sources.*

## 1. Introduction

Air pollution has a big influence on health in many cities in the world [1]. The high concentration of pollutants is very alarming during some circumstances. The pollutants can be of different nature: chemical such as $SO_x$, or $NO_x$; or others such as PM10 (Particulate Matter with an aerodynamic diameter of 10 micrometers). An air pollution crisis occurs when the concentration of air pollutant exceeds maximum exposure levels, thus concentration levels need to be accurately predicted. During these crisis actors (governments and managers) put in place regulations in order to reduce emissions, this is done in several ways including cooperating with the emission source controllers. But, in some circumstances, some of the air pollutants are not controlled: as is the case with those arising from natural sources (e.g. wildfires, volcanic activities, sandstorms), or manmade sources like leaks in industrial processes.

This paper extends our simulation model by including uncontrolled sources of pollutants. We used our simulator to conduct experiments to investigate the effects of cooperation in order to control the crisis taking into account uncontrolled sources.

The air pollution controlling agencies need to cooperate with pollution source controllers (such as factories and road traffic managers) so that they will adapt their emissions and thus help to manage the pollution crisis. This management can be more efficient and realistic if it takes in consideration the natural sources of air pollution and leaks that are not controlled by human actors.

Pollution simulation and decision support tools can help in defining environmental policies in order to preserve the environment and public health during crisis situations. Such models allow decision-makers to assess the effectiveness of theirs policies in the controlling of pollution crisis.

Air pollution has been modelled using different approaches such as: mathematical emission models ([2], [3], [4]), linear models, ANN (Artificial Neural Networks) models ([5], [6]) and hybrid models [7]. Most of them only address the physical and chemical aspects of the phenomenon (concentration and dispersal) and do not take in consideration the human-decision factors and natural sources of air pollution. Air pollution includes a large spatial distribution and it is influenced by the complex interaction of many actors. Human activities (road traffic, industrial and agricultural activities) are considered as the major sources of air pollution. Therefore, it is essential to include human-decision impacts among the simulation process.

MAS (Multi-Agent System) based models are a promising method for modelling pollution related problems [8]. They help us to model the behaviours of humans actor involved the exploitation of natural resources. [9] presents a review of recent MAS models used to study environmental pollution problems. [10] used a MAS approach to investigate the atmospheric pollution emissions resulting from road activities by using a traffic flow simulation linked to emission calculation. [11] used the same approach to study the effect of transport regulations on air pollution emissions.

We used the simulator to run scenarios of air pollution crisis, in order to investigate the possible cooperation between emission source controllers and the air pollution controlling agency. The cooperation can help to predict how the emission sources can participate in managing and reducing the effect of the crisis. The feasibility of our approach is demonstrated by a scenario using data from Annaba, a Mediterranean city in the northeast Algeria.

The paper is organized as follows: Section (2) presents the architecture of the simulator, the dispersion and prediction models and their integration; at the end of this section we define some cooperation strategies. Section (3) presents a description of the simulation scenario using the data from Annaba city. The results of the simulation are presented and discussed in section (4). We end the paper by a conclusion and possible future work.

## 2. Model Approach and Architecture

Our simulation approach is illustrated in figure 1. Agents' actions affect the emission rate of the sources they control. The dispersion model is then used to compute the dispersal; the aggregated value of pollutant concentration is used with weather parameters to forecast the air pollution concentration for *k* hours ahead. According to these forecasts, agents are rewarded or penalised using a regulation formula. We compute how the agent has contributed to the crisis according to the concentration of pollutant that is emitted. Agents then adapt their strategies to earn more rewards and/or reduce penalties.

### 2.1 Dispersion and prediction models

The dispersion model simulates how a pollutant can spread in the atmosphere. The dispersion is computed using the distance from the source, the wind velocity and emission rate. We used a GPD (Gaussian Plum Dispersion) model, which is commonly used in atmospheric dispersion [12]. GPD simulates the dispersal from a point source emission according to (1).

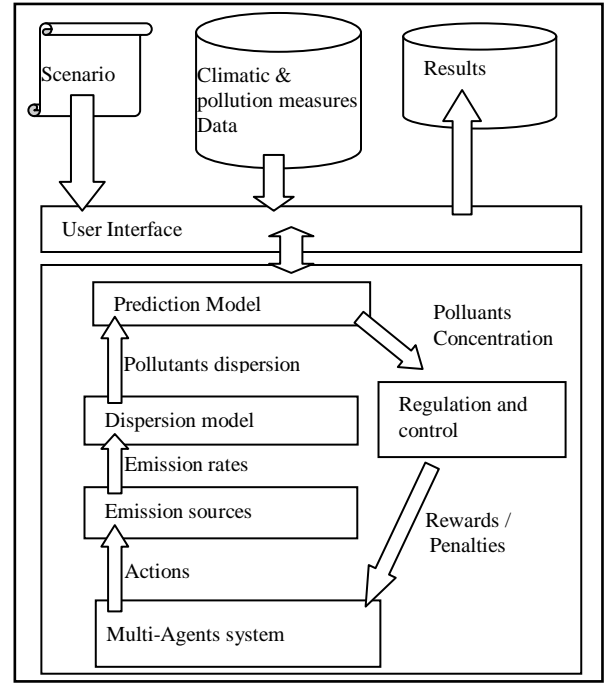

**Figure 1: The components of the simulator.**

$$C(x,y,z,H) = \frac{er_{i,t}}{2\pi U_t\, \sigma_y \sigma_z} * e^{-\frac{y^2}{2*\sigma_y^2}} * \left[ \left(e^{-(\frac{(z-H)^2}{2\sigma_z^2})}\right) + \left(e^{-(\frac{(z+H)^2}{2\sigma_z^2})}\right) \right] \quad (1)$$

At a position (x,y,z) the concentration of the pollutant is calculated according to (1): where

$er_{i,t}$: the emission rate in kilograms per hour of the source i at time step t, and

$U_i$: the wind speed in metres per second at time step t,

$\sigma y, \sigma z$: the standard deviation of the concentration distributions in vertical direction crosswind, these two parameters are chosen according to the stability class 'C' in the **Guifford-Pasquiill** scale, and $H$ is the height of the source from the ground.

The average pollutant concentration from each source is computed. After that, it is passed to a prediction model based on ANN as described in [13].

The ANN predictor uses the aggregated air pollution concentration value from the dispersal models of each source and the four climatic parameters: wind speed, humidity, temperature and rainfall. These parameters greatly influence the pollutant concentration.

The ANN forecasting model is designed to give a forecast of the air pollutant. The combination of the two models (GPD and ANN) can help to introduce an uncertainty aspect caused by the weather conditions.

## 2.2 Management strategy

In our simulator we defined a cooperation strategy for managing the crisis situation. It is based on rewarding/penalising the agents in order to encourage cooperation. This strategy is inspired by game theory where each agent is considered as a player. Every agent has two possible decisions: reducing its emission or increasing it. At each simulation step the agent, based on its previous rewards and the rewards of its neighbours, computes two probabilities: $P$ the probability to reduce emission and $Q$ the probability to increase emission. After that agents are rewarded or penalized according to the number of agents that choose to cooperate. The polluting penalty is computed according to the agent's participation in the pollution.
The algorithm of this strategy is illustrated in figure 2, the reduce emission decision is represented by 0 and the increase decision by 1.

```
Algorithm   Choose Action Emission Agent
while t < MaxTemps do
  if (lastChoices[0] == 0) then
    if ((RPwt < pf_AVG) and (P < Q) and (Q > Ru)) then
      lastChoices[0] = 1;
      sourcesInfo.resumeEmission();
    else
      lastChoices[0] = 0;
      .sourcesInfo.reduceEmission();
    end if
  end if
  if (lastChoices[0] == 1) then
    if ((RPwt < pf_AVG) and (Q < P) and (P > Ru)) then
      lastChoices[0] = 0;
      sourcesInfo.reduceEmission();
    else
      lastChoices[0] = 1;
      sourcesInfo.resumeEmission();
    end if
  end if
end while
```

**Figure 2: The algorithm for Reward/Penalising strategy**

Where: $RPwt$ is the weighted average of the rewards of the agent for the k previous steps.
$PFavg$ is the average of the rewards of its neighbours.
$P$ is the probability of deciding to reduce emissions, whereas $Q$ is the probability of deciding to increase it. These two probabilities are computed according to:

$$\begin{cases} Pc_i(t+1) = Pc_i(t) + 1 - Pc_i(t) * \alpha, & if \ S_i = 0 \ and \ WP_t > 0 \\ Pc_i(t+1) = (1-\alpha) * Pc_i(t), & if \ S_i = 0 \ and \ WP_t \leq 0 \end{cases}$$

(2)

$$\begin{cases} Q_i(t+1) = Q_i(t) + 1 - Q_i(t) * \alpha, & if \ S_i = 1 \ and \ WP_t > 0 \\ Q_i(t+1) = (1-\alpha) * Q(t), & if \ S_i = 1 \ and \ WP_t \leq 0 \end{cases}$$

(3)

Where Wpt is the weighted rewards of the agent $i$ at time step $t$, and $\alpha$ is variable that can have three possible values: 0.015 if the agent does not change its decision in k steps, 0.010 if it changes it once in K times, -0.015 if the agent changes it every simulation step.

## 3. Simulation scenario

The dataset used in this work covers roughly two years, from 2003 to 2004. The data is provided by the local pollution agency network. The data is collected from stations in Annaba city. Annaba is located in the eastern part of the Algerian coast (600 km from Algiers). Annaba city is situated on a vast plain bordered by a mountain chain in the northwest, decreasing in height towards the southwest, and by the Mediterranean Sea on its eastern front. Its bowl shaped topography favours air stagnation and the formation of temperature inversions. Air pollutants are monitored continuously; the concentrations of Particulate Matter (PM10), $SO_x$ and $NO_x$ are measured hourly. The dataset also includes four climatic parameters: Wind Speed (WS), Temperature (T) and relative Humidity (H). Daily measurements about rainfall (RF) were given by the water management agency.. The 2003 dataset was used for training the ANN and the 2004 one was used for validation; this helped us to assess the performance of the model. The pollutant concentration measurements are in microgram/m3. Table 2 presents the statistical properties of the available data for different pollutants and weather parameters, for some parameters data are not available (N/A).

We defined a simulation scenario for the region of Annaba using the parameters in table 3. The simulation scenario contains the values of every simulation parameter grouped in two categories:

(i) Polluting activities and policy parameters: The number of controlled sources of pollution and their maximum emission rates, the number of uncontrolled sources, the goal level of pollutant. This category includes also the cooperation parameters.

(ii) Environment parameters: this category contains the description of the environment such as climatic

parameters and the total number of simulation steps.

**Table 2: Statistical properties of the used dataset.**

| Parameter | 2003 mean | 2004 mean | 2003 STD | 2004 STD | Max value |
|---|---|---|---|---|---|
| PM10 microgram/m$^3$ | 51.70 | 27.76 | 51.66 | 26.38 | 508 |
| NOx | 14.50 | N/A | 25.01 | N/A | 190.0 |
| SO2 | 7.60 | N/A | 14.78 | N/A | 12.8 |
| Wind speed microgram/m$^3$ | 2.65 | 2.12 | 1.78 | 1.27 | 9.6 |
| Humidity (%) | 63.52 | 71.92 | 16.50 | 14.33 | 93.0 |
| Temperature (°C) | 18.96 | 16.82 | 7.76 | 6.30 | 42.1 |
| Rainfall (mm) | N/A | 2.96 | N/A | 9.27 | 73.9 |

**Table 3: Parameters of the simulation scenario.**

| Parameter Name | Value |
|---|---|
| *Polluting activities and Policy parameters* | |
| Number of controlled sources | 240 (80 PM10, 80 NO$_x$ and 80 SO$_x$) |
| Max emission rate | 2000 gram/hour |
| Number of uncontrolled source | 15 (5 PM10, 5 NO$_x$ and 5 SO$_x$) |
| Max emission rate (for uncontrolled) | 5000 gram/hour |
| Goal PM10 level | 70 μ gram/m$^3$ |
| Goal SOx level | 60 μ gram/m$^3$ |
| Goal NO$_x$ | 50 μ gram/m$^3$ |
| Number of memory steps ($K$) | 4 steps |
| Initial proportion of cooperating agents | 0.5 |
| *Environment parameters* | |
| Number of boxes | 20 |
| Temperature at t=0 | 12.7 (°C) |
| Humidity at t=0 | 71.0 % |
| Wind Speed t=0 | 2.4 m/s |
| PM10 at t=0 | 13.0 μ gram/m$^3$ |
| Air Quality at t=0 | 2 ( Good) |
| Total simulation time | 4900 hours |
| Simulation step | 1 step = 2 hours |
| Prediction horizon | 2 hours advance |

## 4. Simulation results

The scenario uses the Annaba dataset. Figure 3 illustrates the user interface of the simulator, which facilitates inspecting the communication between agents and shows the evolution of pollutants concentration during the simulation.

In order to show the impact of uncontrolled pollution during the crisis, we used two scenarios: one with 15 uncontrolled sources (5 for each pollutant) and one without uncontrolled sources. Both scenarios were tested using a cooperation strategy and with no cooperation.

In figure 4, we can see the compare the evolution of PM10 concentration. In the case where the air pollution crisis is managed using a cooperation strategy and with no uncontrolled sources (PM10-Coop-Without leaks), the PM10 concentration is kept under the maximum allowed level.

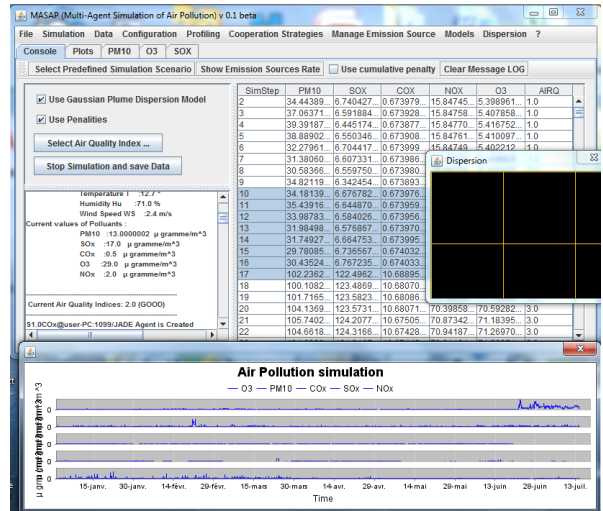

**Figure 3: User interface showing the execution of the simulation scenario.**

Using cooperation when there are uncontrolled sources helps to reduce the PM10 level (PM10-Coop-With leaks), as compared with the case where the cooperation is not used PM10-NC-With-leak. Our simulation results show clearly that cooperation helps in reducing the impact of pollution crises even when some of the emission sources cannot be reduced.

The same evolution is shown in figures 5 and 6 for the case of NO$_X$ and SO$_X$. The results show that cooperation has a big impact on the evolution of pollution.

Interestingly, the simulation results show that even when polluting agents are cooperating at their maximum level, the pollution concentration is still above the recommended level. This means that no matter how much the emission source controllers cooperate they cannot bring the level of pollution below the accepted level. Thus the emission source controllers cannot compensate for the uncontrolled sources (natural or due to leaks). Trying to find a balance between the number of uncontrolled sources and the tolerable emissions is an interesting and valuable question.

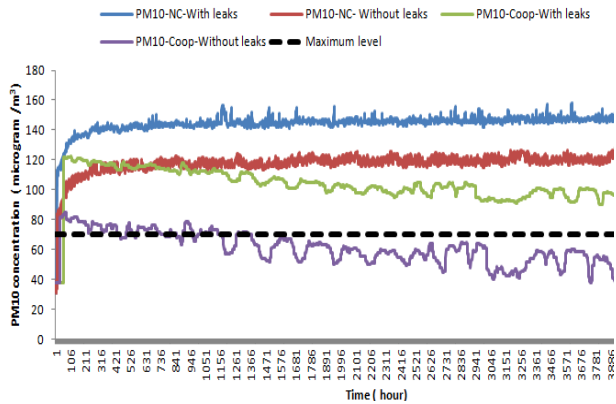

**Figure 4: PM10 concentration for the scenario with and without leak sources.**

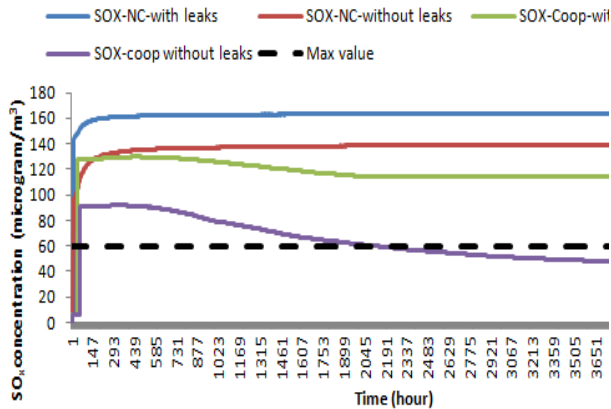

**Figure 5: SO$_x$ concentration for the scenario with and without leak sources.**

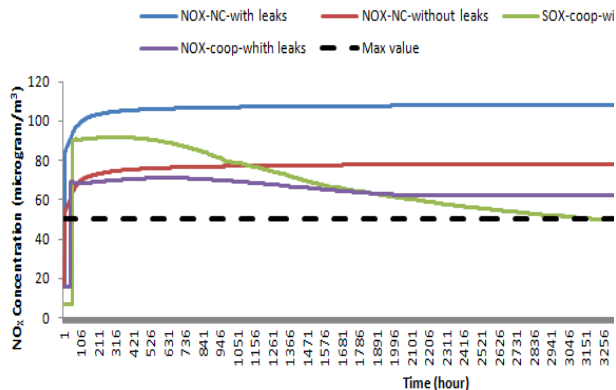

**Figure 6: NO$_x$ concentration for the scenario with and without leak sources.**

## 5. Conclusion

Simulation tool can gives a valuable help to manage pollution related crisis. In our study we showed how a multi-agent based simulation approach can be integrated with dispersion and prediction models in order to simulate an air pollution crisis. Including uncontrolled sources of pollutant is key element for making the simulation more realistic. Since the pollution crises can occurs from sources of different origins (natural, manmade, leaks). Our simulation helped to investigate how cooperation can help controlling or reducing the pollution crisis. The results showed that cooperation between the different actors involved in the management of the crisis, can reduce the effect the pollutant concentration, even if some of the sources are uncontrolled.

The current version of the simulator models point emission sources of PM10, NO$_x$ and SO$_X$. In future versions we aim to include the prediction of others pollutants and also predict air quality. The simulator may also be dotted with a GIS interface, which helps showing geographically the contaminated region. In addition, exploring other cooperation strategies are also among our future plans.

## 6. Acknowledgements

This work was funded by the Algerian Ministry of Higher Education and Scientific Research, PNE 2014/2015 Program.## 7. References

balance between the number of uncontrolled sources and the tolerable emissions is an interesting and valuable question.

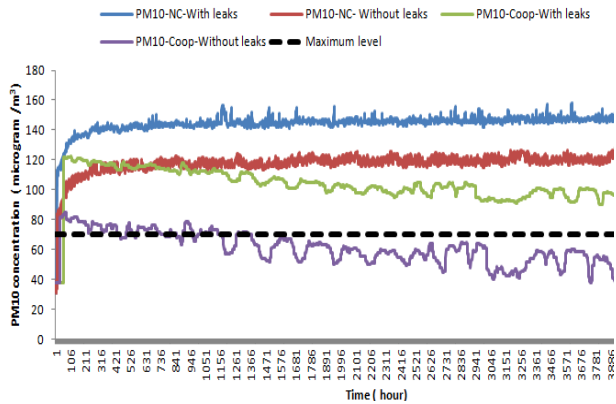

**Figure 4: PM10 concentration for the scenario with and without leak sources.**

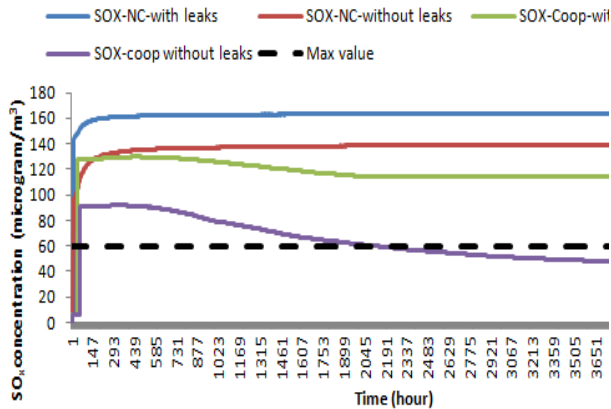

**Figure 5: SO$_x$ concentration for the scenario with and without leak sources.**

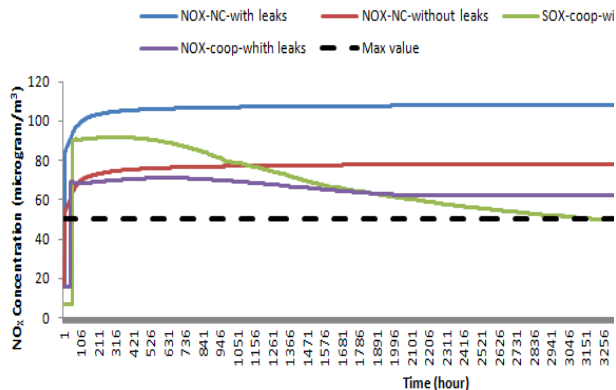

**Figure 6: NO$_x$ concentration for the scenario with and without leak sources.**

## 5. Conclusion

Simulation tool can gives a valuable help to manage pollution related crisis. In our study we showed how a multi-agent based simulation approach can be integrated with dispersion and prediction models in order to simulate an air pollution crisis. Including uncontrolled sources of pollutant is key element for making the simulation more realistic. Since the pollution crises can occurs from sources of different origins (natural, manmade, leaks). Our simulation helped to investigate how cooperation can help controlling or reducing the pollution crisis. The results showed that cooperation between the different actors involved in the management of the crisis, can reduce the effect the pollutant concentration, even if some of the sources are uncontrolled.

The current version of the simulator models point emission sources of PM10, NO$_x$ and SO$_X$. In future versions we aim to include the prediction of others pollutants and also predict air quality. The simulator may also be dotted with a GIS interface, which helps showing geographically the contaminated region. In addition, exploring other cooperation strategies are also among our future plans.

## 6. Acknowledgements

This work was funded by the Algerian Ministry of Higher Education and Scientific Research, PNE 2014/2015 Program.

## 7. References

[1] WHO (World Health Organisation),(2005) Ecosystems and Human Well-being: Health Synthesis. WHO Library Cataloguing-in-Publication Data.
[2] Daly, A., & Zannetti, P. (2007). Air pollution modeling– An overview. Ambient air pollution.
[3] Shuiyuan C., Jianbing L., Beng F., Yuquan J., Ruixia H., A gaussian-box modeling approach for urban air quality management in a northern chinese city—I.model development, Water Air Soil Pollut, 178:37–57, DOI 10.1007/s11270-006-9120-3, 2006.
[4] Lushi, Enkeleida, and John M. Stockie. "An inverse Gaussian plume approach for estimating atmospheric pollutant emissions from multiple point sources." Atmospheric Environment 44.8 (2010). pp. 1097-1107.
[5] Azid, A., Juahir, H., Toriman, M. E., Kamarudin, M. K. A., Saudi, A. S. M., Hasnam, C. N. C., Yamin, M. (2014). Prediction of the Level of Air Pollution Using Principal Component Analysis and Artificial Neural Network Techniques: a Case Study in Malaysia. Water, Air, & Soil Pollution, 225(8), 1-14.
[6] Feng, X., Li, Q., Zhu, Y., Hou, J., Jin, L., & Wang, J. (2015). Artificial neural networks forecasting of PM 2.5 pollution using air mass trajectory based geographic model and wavelet transformation. Atmospheric Environment, 107, 118-128.